\documentclass[12pt,superscriptaddress,aps,prd,preprint]{revtex4-1}
\usepackage{amsmath}
\usepackage{amssymb}
\makeatletter

\newcommand{\bea}{\begin{eqnarray}}
\newcommand{\eea}{\end{eqnarray}}

\begin{document}
\title{ On Non-Commutative Correction of the G\"odel-type Metric}
\author{S. C. Ulhoa}\email[]{sc.ulhoa@gmail.com}
\affiliation{Instituto de F\'{i}sica,
Universidade de Bras\'{i}lia, 70910-900, Bras\'{i}lia, DF,
Brazil.}

\author{A. F. Santos}
\affiliation{Instituto de F\'{\i}sica, Universidade Federal de Mato Grosso,\\
78060-900, Cuiab\'{a}, Mato Grosso, Brazil}
\email{alesandroferreira@fisica.ufmt.br}

\author{R. G. G. Amorim}\email[]{ronniamorim@gmail.com}
\affiliation{Instituto de F\'{i}sica,
Universidade de Bras\'{i}lia, 70910-900, Bras\'{i}lia, DF,
Brazil.} \affiliation{Faculdade Gama, Universidade de Bras\'{i}lia, Setor Leste
(Gama), 72444-240, Bras\'{i}lia-DF, Brazil.}

\begin{abstract}
In this paper, we will study non-commutative corrections in the metric tensor for the G\"{o}del-type universe, a model that has as its main characteristic the possibility of violation of causality, allowing therefore time travel. We also find that the critical radius in such a model, which eventually will determine the time travel possibility, is modified due to the non commutativity of spatial coordinates.
\end{abstract}

\maketitle

\section{Introduction}
Non-commutative geometry arises in the Weyl and Moyal works \cite{weyl}, they studied quantization procedures in phase space. However, Snyder \cite{snyder} was the first one who developed a consistent theory for non-commutative space coordinates which was based on representations of Lie groups. Recently, non-commutative field theories has been extensively studied, thus it is commonly accepted that such a non-commutative feature is directed related to string theory, quantum Hall effect, quantum gravity, Landau levels, M-theory compactification  and Aharonov-Bohm effect \cite{douglas, connes, seiberg, sus, mez}. Usually, the non-commutative spaces can be realized as spaces where coordinates operator $\widehat{x}^{\mu}$ satisfies the commutation relations
\begin{equation}\label{nc1}
[\widehat{x}^{\mu},\widehat{x}^{\nu}]=i\theta^{\mu\nu},
\end{equation}
where $\theta^{\mu\nu}$ is  an anti-symmetric tensor and is of space dimension length squared. The non-commutative models given by Eq.(\ref{nc1}) can be realized by Moyal product or $\star$-product, in which the usual product $f(x)g(x)$ is replaced by the product defined by
\begin{small}
\begin{eqnarray}
f(x) \star g(x) &\equiv& \exp
\left(  {i \over 2} \theta^{\mu\nu}  {\partial \over \partial x^\mu}
{\partial \over \partial y^\nu} \right) f(x) g(y) |_{y \rightarrow x} \nonumber\\
&=& f(x) g(x) + {i\over 2} \theta^{\mu \nu}
\partial_\mu f  \partial_\nu  g + {1 \over 2!}  {\left( i \over 2 \right)^2} \theta^{\mu_1\nu_1} \theta^{\mu_2\nu_2}(\partial_{\mu_1}   \partial_{\mu_2} f )(\partial_{\nu_1}   \partial_{\nu_2} g )+ \cdots\label{starproduct}\nonumber\\
\end{eqnarray}
\end{small}
It interesting to point out that the non-commutativity between spatial and temporal coordinates, $\theta^{0i}\neq0$, can lead to problems with unitarity and causality. Moreover, non-commutativity in space-time coordinates leads to the violation of Lorentz symmetry, a possibility intensively analyzed theoretically and experimentally \cite{car}.

In this work we study the effects of space-time non-commutativity in G\"{o}del-type cosmological model. This cosmological model was proposed by Kurt G\"{o}del in 1949 \cite{Godel}. This is a model that exhibits cylindrical symmetry, matter in rotation and has as main characteristic the possibility of closed type-time curves, the so called CTC's, which lead to violation of causality. A visualization of the G\"{o}del universe with the help of computer graphics is done in \cite{Buser}. In \cite{RebouÃ§as} G\"{o}del-type solution was analyzed in detail, where it found a critical radius which determines the limit to have CTC's or not. Recently this universe model has been widely studied in the literature, for example, in \cite{Santos1, Santos2} was studied the compatibility of this solution with the Chern-Simons modified gravity, in \cite{RebouÃ§as1} was analyzed the $f(R)$ gravity in G\"{o}del-type universe, a study in $f(R,T)$ gravity in the G\"{o}del and G\"{o}del-type models were realized in \cite{Santos3, Ferst}, a study about chronology protection in Horava-Lifshitz gravity in the G\"{o}del-type model was realized in \cite{Petrov}, already \cite{Ulhoa} analyzed the energy-momentum flux in G\"{o}del-type models and in \cite{Ulhoa2} was verified the Rastall gravity in this context. In recent years also emerged a great interest in studying G\"{o}del-type solutions to five-dimensional supergravity \cite{Gauntlett, Brecher}. Therefore, given the interest in this solution we propose to investigate in this article non commutative corrections for this model.

This paper is organized as follows: in section II we present non-commutative corrections of a general metric tensor, in section III we study non-commutative corrections for the G\"{o}del-type metric and the conclusions are displayed in section IV.

\section{Non-Commutative Corrections of the Metric Tensor Components}

In this section we'll deal with non-commutative corrections of the metric tensor components. Thus we start with the following line element in cylindrical coordinates

\bea
ds^2=g_{00}dt^2+g_{02}dtd\phi+g_{11}dr^2+g_{22}d\phi^2+g_{33}dz^2\,,
\eea
where all metric tensor components are functions of $r$ only. This metric represents a stationary rotation around z-axis.

The tetrad field adapted to an observer at rest at spatial infinity, that yields the above metric, is

\bea
e^a\,_\mu=\left(
            \begin{array}{cccc}
              A & 0 & H & 0 \\
              0 & B\cos\phi & -D\sin\phi & 0 \\
              0 & B\sin\phi & D\cos\phi & 0 \\
              0 & 0 & 0 & C \\
            \end{array}
          \right)\,,
\eea
where
\begin{eqnarray}
A&=&\sqrt{-g_{00}}\,,\nonumber\\
H&=&-\frac{g_{02}}{A}\,,\nonumber\\
B&=&\sqrt{g_{11}}\,,\nonumber\\
D&=&\frac{(g_{02}^2-g_{00}g_{22})^{1/2}}{A}\,,\nonumber\\
C&=&\sqrt{g_{33}}\,.
\end{eqnarray}

In order to induce non-commutative corrections on the metric tensor $g_{\mu\nu}\rightarrow\tilde{g}_{\mu\nu}$, we substitute the normal product by the Moyal product in the definition $g_{\mu\nu}=e^a\,_\mu e_{a\nu}$. Thus we get

$$\tilde{g}_{\mu\nu}=\frac{1}{2}\left(e^a\,_\mu\star
e_{a\nu}+e^a\,_\nu\star e_{a\mu}\right)\,,$$ in such a relation we also had to symmetrize the definition.
The tetrad field transforms like $e'^a\,_\mu=\Lambda^a\,_b e^b\,_\mu$, under SO(3,1) transformations, where $\Lambda^a\,_b$ is the
Lorentz matrix.  Thus $\tilde{g}_{\mu\nu}$ represents a deformation in the diffeomorphic group due to the non-commutativity between coordinates. It has been showed that corrections in the
metric tensor appears up to second order in the non-commutative
parameter
$\theta^{\mu\nu}$~\cite{Mukherjee:2006nd,Chaichian2008573}.

If we choose non-commutativity between $r$ and $\phi$, which is realized by
\bea
\theta^{\mu\nu}=\left(
                  \begin{array}{cccc}
                    0 & 0 & 0 & 0 \\
                    0 & 0 & \alpha & 0 \\
                    0 & -\alpha & 0 & 0 \\
                    0 & 0 & 0 & 0 \\
                  \end{array}
                \right)\,,
\eea
then, with the help of the definition $\Delta g_{\mu\nu}=\tilde{g}_{\mu\nu}-g_{\mu\nu}$, we find the following non-vanishing components

\begin{eqnarray}
\Delta g_{11}&=&\frac{\alpha^2}{8}\left(\frac{\partial^2g_{11}}{\partial r^2}\right)\,,\nonumber\\
\Delta g_{22}&=&\frac{\alpha^2}{8g_{00}^3}\Biggl\{\left[\frac{\partial\left(g_{00}^2\right)}{\partial r}\right]\left[\frac{\partial\left(g_{02}^2\right)}{\partial r}\right]+g_{00}^3\left(\frac{\partial^2g_{22}}{\partial r^2}\right)+g_{00}g_{02}^2\left(\frac{\partial^2g_{00}}{\partial r^2}\right)\nonumber\\
&-&2g_{00}^2\left[\left(\frac{\partial g_{02}}{\partial r}\right)^2+g_{02}\left(\frac{\partial^2 g_{02}}{\partial r^2}\right)\right]-g_{02}^2\left(\frac{\partial g_{00}}{\partial r}\right)^2\Biggr\}\,.\label{cor}
\end{eqnarray}
In the next section we apply this to the G\"{o}del-type Universe.

\section{G\"odel-Type Universe}

Here we will study non-commutative corrections for the G\"{o}del-type metric that is given by \cite{RebouÃ§as}
\bea
ds^2=-\left[dt+H(r)d\phi\right]^2+dr^2+D(r)^2d\phi^2+dz^2,
\eea
with
\bea
H(r)&=&\frac{4\omega}{m^2}\,senh^2\left(\frac{mr}{2}\right),\label{H}\\
D(r)&=&\frac{1}{m}\,senh(mr),\label{D}
\eea
where $m$ and $\omega$ are parameters that characterize all G\"{o}del-type metrics.

In this case, simplifying expressions (\ref{cor}),  the new components of the metric tensor due to the non-commutativity of space-time, up to second order in the non-commutativity parameter, are
\bea
\tilde{g}_{00}&=&g_{00},\nonumber\\
\tilde{g}_{02}&=&g_{02},\nonumber\\
\tilde{g}_{11}&=&g_{11},\nonumber\\
\tilde{g}_{22}&=&g_{22}+\frac{1}{4}\alpha^2\left[D\left(\frac{d^2D}{dr^2}\right)+\left(\frac{dD}{dr}\right)^2\right],\,\nonumber\\
\tilde{g}_{33}&=&g_{33}.
\eea

Thus, we note that the corrections of the non-commutative affects only the component $g_{22}$. Using the equations (\ref{H}) and (\ref{D}) we can rewrite this component as
\begin{equation}
\tilde{g}_{22}=G(r)+\frac{1}{8}\alpha^2\,\cosh{(2mr)}\,,
\end{equation}
where $G(r)=D^2(r)-H^2(r)$.

So far we have seen that there are non-commutative corrections to the metric G\"{o}del-type as shown in the above equation. As discussed in the introduction, the presence of CTC's is one of the main features of the  G\"{o}del-type solutions. Here we can analyze whether the non-commutativity of spacetime affects these closed time-like curves that lead to violation of causality analyzing the critical radius. First, let us recall how to determine the critical radius for the commutative case. In this case, the existence  of CTC's is related to the behavior of the function $G(r)$ in the interval $r_1<r<r_2$ \cite{RebouÃ§as}. If this function becomes negative in this interval, the curve defined by $t,r,z=constante$ is a closed timelike that leads to violation of causality. In this case we find
\bea
sinh\left(\frac{mr}{2}\right)=\frac{m}{\sqrt{4\omega^2-m^2}},
\eea
and the critical radius becomes
\bea
r_c=\frac{2}{m}sinh^{-1}\left(\frac{m}{\sqrt{4\omega^2-m^2}}\right).\label{r1}
\eea
We recover the G\"{o}del metric when $m^2=2\omega^2$.

Now to check the influence of non-commutative corrections to the critical radius should analyze the behavior of function $\tilde{G}(r)$ defined by
\bea
\tilde{G}(r)&=&G(r)+\frac{1}{8}\alpha^2\,\cosh{(2mr)},\nonumber\\
&=&\frac{4}{m^2}sinh^2\left(\frac{mr}{2}\right)\left[\left(1-\frac{4\omega^2}{m^2}\right)sinh^2\left(\frac{mr}{2}\right)+1\right]+\frac{1}{8}\alpha^2\,\cosh{(2mr)}.
\eea
And thus we find that critical radius which determines the existence of CTC in the presence of non-commutative corrections is
\bea
\tilde{r}_c=\frac{2}{m}sech^{-1}\left[\frac{2\sqrt{\alpha^2m^4+4m^2-16\omega^2}}{\sqrt{2\alpha^2m^4+8m^2-64\omega^2-[2m^4(\alpha^4m^4+4\alpha^2(3m^2+4\omega^2)+32)]^{1/2}}
}\right].\label{r2}
\eea
Therefore, the critical radius suffers corrections due to  non-commutativity of space.

At this point, we will analyze if the equation (\ref{r2}) is reduced to the usual value, equation (\ref{r1}), when we take $\alpha=0$. In this case we obtain
\bea
\tilde{r}_c=\frac{2}{m}cosh^{-1}\left(\frac{2\omega}{\sqrt{4\omega^2-m^2}}\right),
\eea
where we use the relation $sech^{-1}(x)=cosh^{-1}\left(\frac{1}{x}\right)$. Using that
\bea
sinh[cosh^{-1}(x)]=\sqrt{\frac{x-1}{x+1}}(x+1)=\sqrt{x^2-1},
\eea
we stay with
\bea
sinh\left(\frac{m\tilde{r}}{2}\right)=\frac{m}{\sqrt{4\omega^2-m^2}},
\eea
therefore, $\tilde{r}_c=r_c$, and thus we show that when $\alpha=0$ we recover the result obtained in equation (\ref{r1}).

\section{Conclusion}

In this work we studied the influence of non-commutative corrections of metric tensor into the features of G\"{o}del-type Universe. We found generic corrections for the metric tensor components of a cylindrical rotating geometry once we assumed non-commutativity between $r$ and $\phi$ which is set by the choice $\theta^{12}=\alpha$ as the only non-vanishing component of $\theta^{\mu\nu}$. Then we applied our expressions to G\"odel-type metric and established that only the component $g_{22}$ of the metric was changed by corrections until the second order in the parameter of non-commutativity. We also show that the non-commutativity modifies the critical radius that determines the limit for the existence of closed timelike curves (CTC's) in G\"{o}del-type universe.

\begin{acknowledgements}
This work was partially supported by Conselho Nacional de Desenvolvimento Cient\'{\i}fico e Tecnol\'{o}gico (CNPq) and Coordena\c{c}\~ao de Aperfei\c{c}oamento de Pessoal de N\'ivel Superior (CAPES). A. F. S. has been suported by the CNPq project 476166/2013-6.
\end{acknowledgements}


\begin{thebibliography}{99}%
\makeatletter
\providecommand \@ifxundefined [1]{%
 \@ifx{#1\undefined}
}%
\providecommand \@ifnum [1]{%
 \ifnum #1\expandafter \@firstoftwo
 \else \expandafter \@secondoftwo
 \fi
}%
\providecommand \@ifx [1]{%
 \ifx #1\expandafter \@firstoftwo
 \else \expandafter \@secondoftwo
 \fi
}%
\providecommand \natexlab [1]{#1}%
\providecommand \enquote  [1]{``#1''}%
\providecommand \bibnamefont  [1]{#1}%
\providecommand \bibfnamefont [1]{#1}%
\providecommand \citenamefont [1]{#1}%
\providecommand \href@noop [0]{\@secondoftwo}%
\providecommand \href [0]{\begingroup \@sanitize@url \@href}%
\providecommand \@href[1]{\@@startlink{#1}\@@href}%
\providecommand \@@href[1]{\endgroup#1\@@endlink}%
\providecommand \@sanitize@url [0]{\catcode `\\12\catcode `\$12\catcode
  `\&12\catcode `\#12\catcode `\^12\catcode `\_12\catcode `\%12\relax}%
\providecommand \@@startlink[1]{}%
\providecommand \@@endlink[0]{}%
\providecommand \url  [0]{\begingroup\@sanitize@url \@url }%
\providecommand \@url [1]{\endgroup\@href {#1}{\urlprefix }}%
\providecommand \urlprefix  [0]{URL }%
\providecommand \Eprint [0]{\href }%
\providecommand \doibase [0]{http://dx.doi.org/}%
\providecommand \selectlanguage [0]{\@gobble}%
\providecommand \bibinfo  [0]{\@secondoftwo}%
\providecommand \bibfield  [0]{\@secondoftwo}%
\providecommand \translation [1]{[#1]}%
\providecommand \BibitemOpen [0]{}%
\providecommand \bibitemStop [0]{}%
\providecommand \bibitemNoStop [0]{.\EOS\space}%
\providecommand \EOS [0]{\spacefactor3000\relax}%
\providecommand \BibitemShut  [1]{\csname bibitem#1\endcsname}%
\let\auto@bib@innerbib\@empty

                                                                                               %
\bibitem{weyl}H. Weyl, Z. Phys. \textbf{46} (1927) 1.
\bibitem{snyder} H.S. Snyder, Phys. Rev. \textbf{71} (1947) 38.
\bibitem{douglas} M.R. Douglas, N.A. Nekrasov, Rev. Mod. Phys. \textbf{73} (2001) 977.
\bibitem{connes} A. Connes, M.R. Douglas, A. Schwarz, JHEP \textbf{9802} (1998) 003.
\bibitem{seiberg} N. Seiberg, E. Witten, JHEP \textbf{9909} (1999) 032.
\bibitem{sus} L. Susskind, hep-th/0101029.
\bibitem{mez} L. Mezincescu, hep-th/0007046.
\bibitem{car} S.M. Carroll, J.A. Harvey, V.A. Kostelecky, C.D. Lane, T. Okamoto, Phys. Rev. Lett. \textbf{87} (2001) 141610.
\bibitem{Godel} K. G\"{o}del, Rev. Mod. Phys. {\bf 21}, 447 (1949).
\bibitem{Buser} M. Buser, E. Kajari, W. P. Schleich, New J. Phys. {\bf 15}, 013063 (2013), arXiv:1303.4651.
\bibitem [{\citenamefont {Mukherjee}\ and\ \citenamefont
  {Saha}(2006)}]{Mukherjee:2006nd}%
  \BibitemOpen
  \bibfield  {author} {\bibinfo {author} {\bibfnamefont {P.}~\bibnamefont
  {Mukherjee}}\ and\ \bibinfo {author} {\bibfnamefont {A.}~\bibnamefont
  {Saha}},\ }\href {\doibase 10.1103/PhysRevD.74.027702} {\bibfield  {journal}
  {\bibinfo  {journal} {Phys.Rev.}\ }\textbf {\bibinfo {volume} {D74}},\
  \bibinfo {pages} {027702} (\bibinfo {year} {2006})},\ \Eprint
  {http://arxiv.org/abs/hep-th/0605287} {arXiv:hep-th/0605287 [hep-th]}
  \BibitemShut {NoStop}%
\bibitem [{\citenamefont {Chaichian}\ \emph {et~al.}(2008)\citenamefont
  {Chaichian}, \citenamefont {Tureanu},\ and\ \citenamefont
  {Zet}}]{Chaichian2008573}%
  \BibitemOpen
  \bibfield  {author} {\bibinfo {author} {\bibfnamefont {M.}~\bibnamefont
  {Chaichian}}, \bibinfo {author} {\bibfnamefont {A.}~\bibnamefont {Tureanu}},
  \ and\ \bibinfo {author} {\bibfnamefont {G.}~\bibnamefont {Zet}},\ }\href
  {\doibase 10.1016/j.physletb.2008.01.029} {\bibfield  {journal} {\bibinfo
  {journal} {Physics Letters B}\ }\textbf {\bibinfo {volume} {660}},\ \bibinfo
  {pages} {573 } (\bibinfo {year} {2008})}\BibitemShut {NoStop}%
\bibitem{RebouÃ§as} M. Rebou\c{c}as and J. Tiomno, Phys. Rev. D {\bf 28}, 1251 (1983);
\bibitem{Santos1} C. Furtado, T. Mariz, J. R. Nascimento, A. Yu. Petrov and A. F. Santos, Phys. Rev. D {\bf 79}, 124039 (2009).
\bibitem{Santos2}  C. Furtado, J. R. Nascimento, A. Yu. Petrov and A. F. Santos, Phys. Lett. B {\bf 693}, 494 (2010).
\bibitem{RebouÃ§as1} M. J. Rebou\c{c}as and J. Santos, Phys. Rev. D {\bf 80}, 063009 (2009).
\bibitem{Santos3} A. F. Santos, Mod. Phys. Lett. A {\bf 28}, 1350141, arXiv:1308.3503.
\bibitem{Ferst} C. J. Ferst and A. F. Santos, arXiv: arXiv:1411.1002.
\bibitem{Petrov} J. B. Fonseca-Neto, A. Yu. Petrov, M. J. Rebou\c{c}as, Phys. Lett. B {\bf 725}, 412 (2013),  arXiv:1304.4675.
\bibitem{Ulhoa} S. C. Ulhoa, A. F. Santos, R. G. G. Amorim, Mod. Phys. Lett. A {\bf 28}, 1350039 (2013),  arXiv:1303.0713.
\bibitem{Ulhoa2} A. F. Santos, S. C. Ulhoa, arXiv: 1407.4322.
\bibitem{Brecher} D. Brecher, U. H. Danielsson, J. P. Gregory, M. E. Olsson, J. High Energy Phys. {\bf 0311}, 033 (2003).
\bibitem{Gauntlett} J. P. Gauntlett, J. B. Gutowski, C. M. Hull, S. Pakis, and H. S. Reall, Class. Quant. Grav. {\bf 20}, 4587 (2003), hep-th/0209114.
\end{thebibliography}
\end{document}